\documentclass[10pt,twocolumn,aps,prl,superscriptaddress,footinbib]{revtex4-2}
\usepackage{amsfonts}
\usepackage{amsmath}
\usepackage{amssymb}
\usepackage{color}
\usepackage{graphicx}
\usepackage{xcolor}
\usepackage{bm}
\usepackage[colorlinks,bookmarks=false,allcolors=blue,hyperfootnotes=true]{hyperref}

\begin{document}

\title{Paramagnetic limit of spin-triplet superconductors}

\date{July 17, 2024}

\author{Thomas Bernat}
\author{Julia S. Meyer}
\author{Manuel Houzet}
\affiliation{Univ. Grenoble Alpes, CEA, Grenoble INP, IRIG, PHELIQS, 38000 Grenoble,
France}

\begin{abstract}
We study the phase diagram of spin-triplet superconductors, considering the effect of  the external magnetic field on the electrons' spins. For a given symmetry of the order parameter and a generic orientation of the field, we find that the paramagnetic limit for superconductivity diverges at low temperatures. Furthermore, we identify a range of temperatures where the transition between normal and superconducting phases becomes of the first order. When two tricritical points exist along the transition line, a first order phase transition between two superconducting phases may develop in vicinity of the tricritical point with lower temperature. We discuss the implications of our findings for the anisotropy of the upper critical field in UPt$_3$, a candidate material  for triplet superconductivity, when both the paramagnetic and orbital effects are taken into account.
\end{abstract}

\maketitle

{\it Introduction.--} The temperature dependence of the upper critical field provides a wealth of information about the properties of a superconducting material. Usually, the dominant mechanism for the suppression of superconductivity is the orbital effect \cite{Maki1964a,*Maki1964b,Helfand1966}. The paramagnetic effect may become significant when the electrons' orbital motion is quenched, either by the reduced dimensionality of the material or by a large effective mass, such as in heavy-fermion systems. In that case, the upper critical field provides direct information about the spin content of Cooper pairs. Indeed, the competition between Cooper pairing and Pauli paramagnetism yields a fundamental upper limit for the stability of spin-singlet pairs~\cite{Clogston1962, Chandrasekhar1962,Sarma1963,Sugiyama1993}. The Pauli limit for an s-wave superconductor is given as $H_P=\Delta_0/(\sqrt 2\,\mu)$, where $\Delta_0$ is the zero-temperature order parameter and $\mu$ is the effective magnetic moment. The same mechanism applies for triplet pairs with opposite spins. By contrast, there is no such limit for triplet pairs with equal spins~\cite{Scharnberg1980,Lukyanchuk1987,Mineev2010}. Thus, an upper critical field in excess of the paramagnetic limit is usually taken as a strong indication of triplet superconductivity, see, e.g., Refs.~\cite{Nakamura2019, Wang2021}. In this work, we study the unusual angular dependence of the critical field due to the interplay of the three spin channels in spin-triplet superconductors.

Materials that are suspected to harbor triplet superconductivity are usually made of heavy atoms. Thus, their conduction band is subject to strong spin-orbit coupling. Thereby the spin content of Cooper pairs is tied to crystal axes. Our first finding is that, for a generic structure of the order parameter, the paramagnetic limit is absent at zero temperature for {\it any} direction of the field.  Namely, the paramagnetic limit is only present in the singular case of a triplet order parameter consisting of pairs with zero spin projection along a given crystal axis {\it and} for a magnetic field pointing along that axis. In all other cases, a fraction of equal-spin pairs with respect to the field axis is present and may condense, leading to a divergence of the upper critical field (up to the orbital limit). Our second finding is that a finite temperature restores the pair-breaking effect of pairs with zero spin projection, yielding a finite paramagnetic limit, unless the condensate contains only equal-spin pairs with respect to the field direction. This leads to distinctive features in the temperature dependence and anisotropy of the upper critical field in comparison with those arising from the orbital effect. Furthermore, we find that specific structures of the order parameter may allow for two successive changes of the transition order along the upper critical line between normal and superconducting phases in some range of field orientation. By analogy with the singlet case~\cite{Sarma1963}, the appearance of a tricritical point as the temperature decreases below the superconducting critical temperature is attributed to the triplet pairs with opposite spins. This tricritical point is then accompanied by another one at low temperature, in accordance with our observation that equal-spin pairs are responsible for the second order transition at low temperature. Our third finding is that, when two tricritical points arise along the upper critical line, the competition between triplet pairings with equal and opposite spins may result in a first order transition developing from the tricritical point with lower temperature within the superconducting region of the phase diagram. The corresponding critical line either ends at a critical point, or extends down to  zero temperature.

In the following, we develop a model that allows us to quantify these findings by applying the quasiclassical theory of superconductivity. We show that our first two findings are general for any structure of the order parameter in the regime of strong spin-orbit coupling. We then consider specific examples of triplet order parameters compatible with the crystalline symmetry of the lattice in order to study the full phase diagram.

{\it Model.--} 
To establish the superconducting phase diagram, we start with the free energy functional of a unitary spin-triplet superconductor at temperature $T$ \cite{Mineev1999},
\begin{equation}
F_S = \sum_{\pm,|\xi_{\mathbf k}|<{\cal E}}\left[ \dfrac{\xi_\mathbf{k}}{2} - T\ln\left(2\cosh\dfrac{E_{\mathbf{k},\pm}}{2T}\right) \right] + \dfrac{\nu_0 |\Delta|^2}{\lambda}
	\label{eq:free_energy_integral}
\end{equation}
in units with  $\hbar=k_B=1$. Here, $E_{\mathbf{k},\pm}~=~[\xi_\mathbf{k}^2 + |\mathbf{d_k}|^2 + h^2 \pm 2|h|(\xi_\mathbf{k}^2 + |\mathbf{d}_\mathbf{k}\cdot\hat{\mathbf{h}}|^2)^{1/2}]^{1/2}$ are the positive eigenvalues of the particle-hole symmetric Bogoliubov-de Gennes Hamiltonian 
\begin{equation}
\check{H}_{\mathbf{k}} = \xi_\mathbf{k}\tau_z - \mathbf{h}\cdot \bm{\sigma} + \check{\Delta}_\mathbf{k},
\label{eq:BdG}
\end{equation}
where $\xi_{\mathbf k}$ is the excitation energy of a quasiparticle with momentum $\mathbf k$ in the normal state, $\mathbf{h}=h\hat{\mathbf h}$ is the Zeeman field, and the gap matrix
\begin{equation}
\check{\Delta}_\mathbf{k} = \begin{pmatrix}
0 & \hat{\Delta}_\mathbf{k} \\
\hat{\Delta}_\mathbf{k}^\dagger & 0
\end{pmatrix}\quad \mathrm{with}\quad \hat{\Delta}_\mathbf{k}=\mathbf{d}_\mathbf{k} \cdot \bm{\sigma}
\end{equation}
is related with the spin-triplet order parameter $\mathbf{d_k}=\Delta \bm{\psi}_\mathbf{k}$ with amplitude $\Delta$ and normalized vector function $\bm{\psi}_{\mathbf k}$, tied to crystal axes and with Fermi surface average $\langle |\bm{\psi}_{\mathbf k}|^2 \rangle_\mathbf{k} = 1$. For simplicity, we neglected multiband effects and considered a unitary triplet phase, such that $ \bm{\psi}_\mathbf{k} \times \bm{\psi}^*_\mathbf{k}=0$. The mean-field approximation, which yields the order parameter in Eq.~\eqref{eq:BdG}, involves the pairing amplitude $\lambda$ acting in an energy window $\cal E$ around the Fermi level, such that the superconducting critical temperature in the weak-coupling regime ($\lambda\ll 1$), is $T_c\simeq 1.13\,{\cal E} e^{-1/\lambda}$; $\nu_0$ is the normal density of states. Furthermore, $\bm{\sigma}$ is the vector of Pauli matrices in spin space, and $\tau_z$ is a Pauli matrix in Nambu space.

It is convenient to subtract the free energy in the normal state from Eq.~\eqref{eq:free_energy_integral}. Taking the continuum limit, the sum over momenta can be decomposed as $\sum_\mathbf{k} \dots = \nu_0 \int d\xi_\mathbf{k} \langle \dots \rangle_\mathbf{k}$. 
Deforming the $\xi_\mathbf{k}$-integration contour from real to imaginary axis, we find the  free energy difference functional \cite{SuppMat}, 
\begin{eqnarray}
	F = 2\nu_0\pi T \sum_{|\omega|<\cal E} \left\langle |\omega| - \textrm{Re} \; \Omega_\mathbf{k} \right\rangle_\mathbf{k} + \dfrac{\nu_0|\Delta|^2}{\lambda}.
	\label{eq:free_energy_matsubara}
\end{eqnarray}
Here, $\omega=(2n+1)\pi T$ ($n$ integer) are Matsubara frequencies at temperature $T$ and $\Omega_\mathbf{k} = [\omega^2+\Delta^2|\mathbf{\psi_k}|^2-h^2 + 2ih(\omega^2+\Delta^2|\mathbf{\psi_k}\times\hat{\mathbf{h}}|^2)^{1/2}]^{1/2}$. In particular, the saddle point of Eq.~\eqref{eq:free_energy_matsubara} yields the self-consistent gap equation,
\begin{equation}
	\dfrac{1}{\lambda} = \pi T \sum_{|\omega|<\cal E} \left\langle |\bm{\psi}_{\mathbf k}|^2\textrm{Re} \; \Omega_\mathbf{k}^{-1} + \dfrac{|\bm{\psi}_{\mathbf k}\times\mathbf{h}|^2}{|\Omega_\mathbf{k}|^2 \textrm{Re} \; \Omega_\mathbf{k}} \right\rangle_\mathbf{k}.
	\label{eq:gap_triplet_explicit}
\end{equation}
Below we find the free energy of a given phase by inserting the solution $\Delta$ of Eq.~\eqref{eq:gap_triplet_explicit} into Eq.~\eqref{eq:free_energy_matsubara}.

{\it Second order transition.--} 
The solution of Eq.~\eqref{eq:gap_triplet_explicit} at $\Delta=0$ or, equivalently,
\begin{eqnarray}
	\ln\dfrac{T}{T_c} = \langle|\bm{\psi}_{\mathbf k}\cdot\hat{\mathbf{h}}|^2\rangle_\mathbf{k} \left[ \psi\left(\frac{1}{2}\right) - \mathrm{Re} \; \psi\left(\frac{1}{2}+\frac{ih}{2\pi T}\right) \right]
	\label{eq:gap_D=0_triplet_psi},
\end{eqnarray}
determines the upper critical line corresponding to a second order transition from normal to superconducting phase. Here, $\psi$ is the digamma function. 

The prefactor in the r.h.s.~of Eq.~\eqref{eq:gap_D=0_triplet_psi} radically changes the behavior of the upper critical line, in comparison with the singlet case where this prefactor is equal to 1. To show this, let us introduce the angle $\bar\theta$ such that $\cos^2\bar\theta= \langle|\bm{\psi}_\mathbf{k}\cdot\hat{\mathbf{h}}|^2\rangle_\mathbf{k}$. If ${\bm{\psi}}_{\mathbf{k}}$ keeps a constant direction along the Fermi surface, then $\bar\theta$ is simply the angle between $ \mathbf{h} $ and ${\bm{\psi}}_{\mathbf{k}}$. In general, however, $\bar\theta$ depends both on the direction of the field and the specific structure of the order parameter encoded by the function $\bm{\psi}_\mathbf{k}$. Note that the minimal value $\bar\theta=0$ can only be reached when ${\bm{\psi}}_{\mathbf{k}}$ keeps a constant direction, in which case it corresponds to $\bm{\psi}_\mathbf{k}$ parallel to $\mathbf{h}$, whereas the maximal value $\bar\theta=\pi/2$ can only be reached when ${\bm{\psi}}_{\mathbf{k}}$ spans at most a plane, where it corresponds to $\mathbf{h}$ perpendicular to this plane. The temperature dependence of the upper critical line is easily explained at $\bar\theta=0$ or $\pi/2$~\cite{Scharnberg1980,Lukyanchuk1987,Choi1991}. At $\bar\theta=0$, the superconducting phase is made of pairs with opposite spins and the upper critical line coincides with the one in the singlet case. At $\bar\theta=\pi/2$, the superconducting phase is made of pairs with equal spins and there is no paramagnetic limit: $T_c$ is independent of $h$.

Importantly, there is no symmetry requirement that can fix the direction of ${\bm{\psi}}_{\mathbf{k}}$ for all $\mathbf k$ on the Fermi surface (in accordance with ``Blount theorem'' \cite{Yip1993}). Therefore, $\bar\theta$ is generically different from 0 or $\pi/2$ at {\it any} orientation of the Zeeman field. For instance, a basis function of the one-dimensional representation $A_{1u}$ in the hexagonal symmetry class $D_{6h}$, relevant for UPt$_3$, is $\bm{\psi}^\eta_{\mathbf k}\propto (\eta \hat k_x,\eta \hat k_y, \hat k_z)$ with {\it a priori} finite real coefficient $\eta$ determined by the microscopics of the material \cite{Yip1993}. Assuming a spherical Fermi surface, we find $\cos^2\bar\theta=(\cos^2\phi+\eta^2\sin^2\phi)/(1+2\eta^2)$ where $\phi$ is the tilt angle between $\mathbf h$ and $\hat{\mathbf z}$-axis. In particular, for $\eta<1$, one may realize angles $\bar\theta\in[\arccos(1/\sqrt{1+2\eta^2}),\arccos(\eta/\sqrt{1+2\eta^2})]$. Then, an important consequence of Eq.~\eqref{eq:gap_D=0_triplet_psi} at $\bar\theta\neq0,\pi/2$ is that all triplet phases are paramagnetically limited at finite temperature, and the upper critical field diverges as a power-law as $T$ decreases,
\begin{equation}
	h_c =\dfrac{\Delta_0}{2}\left(\dfrac{T_c}{T}\right)^{\tan^2\bar\theta}, \qquad  T\ll T_c.
	\label{eq:hc_divergence}
\end{equation}
Here, $\Delta_0 \simeq 1.76 T_c$ is the BCS gap at $T~=~0$, which differs from the gap at $T=h=0$, namely $\Delta(0)=~\Delta_0 \exp \langle -|\bm{\psi}_{\mathbf k}|^2\ln|\bm{\psi}_{\mathbf k}|\rangle_\mathbf{k}$. Note that the above considerations hold in the nonunitary case as well.

The temperature dependence of the upper critical field at the second order phase transition is shown in Fig.~\ref{fig:hc(T)_2nd_(0,0,kz)}. Like in the singlet case \cite{Sarma1963,Sugiyama1993}, the non-monotony of the transition line at $\bar\theta < 0.14 \pi$ (reachable when $\eta<0.33$ for the order parameter discussed above) suggests a change of the transition order, as we examine below.

\begin{figure}
	\centering
	\includegraphics[]{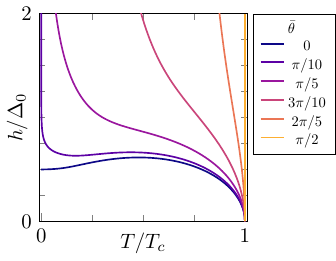}
	\caption{Temperature dependence of the upper critical field at the second order phase transition for different angles $\bar\theta$ characterizing the tilt between $\mathbf{h}$ and $\bm{\psi}_\mathbf{k}$.}
	\label{fig:hc(T)_2nd_(0,0,kz)}
\end{figure}

{\it First order transitions.--} 
The change of the transition order can be inferred from an increased number of solutions of the gap equation~\eqref{eq:gap_triplet_explicit}. 

We first consider the case of a $d$-vector with a constant direction, and a magnetic field $\mathbf{H} = \mathbf{h}/\mu $ along that direction ($\bar\theta=0$) at $T=0$. Then, the gap equation,
\begin{eqnarray}
	\left\langle |\bm{\psi}_{\mathbf k}|^2 \mathrm{Re}\ln\dfrac{\Delta_0}{\sqrt{h^2 - \Delta^2 |\bm{\psi}_\mathbf{k}|^2} + |h|}  \right\rangle_\mathbf{k} = 0,
	\label{eq:gap_theta=0}
\end{eqnarray}
admits for two finite solutions if $\Delta_0/2 < h < \Delta_h^{\rm max} < \Delta_0$, like in the singlet case, though $\Delta_h^{\rm max}$ now depends on the choice of the order parameter. It is illustrated by the dark blue line in Fig.~\ref{fig:D(H)_(0,0,kz)}(a) for the order parameter $\bm{\psi}_\mathbf{k} \propto (0, 0, \hat k_z)$, where $\Delta_h^{\rm max} = 0.75\Delta_0$. Thus, a first order transition occurs when the free energy difference between the normal phase ($\Delta=0$) and the superconducting phase (larger $\Delta$) vanishes. (The solution with intermediate $\Delta$ is unstable.) The corresponding critical field is $h_c=0.66 \Delta_0$.

At $\bar\theta\neq0$ and zero temperature, $\Delta(0)$ is the only solution at $h=0$. There is also a unique solution at $h\gg \Delta$ \cite{SuppMat},
\begin{eqnarray}
	\Delta(h)= \Delta_0 \left(\frac{\Delta_0}{2h}\right)^{\cot^2\bar\theta} e^{\left\langle -|\bm{\psi}_{\mathbf k}\times\hat{\mathbf{h}}|^2\ln|\bm{\psi}_{\mathbf k}\times\hat{\mathbf{h}}| \right\rangle_\mathbf{k}/\sin^2\bar\theta},
	\label{eq:h(D)_h>D}
\end{eqnarray}
which reduces to
\begin{eqnarray}
	\Delta(h) = \frac{\Delta(0)}{\sin \bar\theta} \left(\frac{\Delta_0}{2h}\right)^{\cot^2\bar\theta},
\end{eqnarray}
for a $d$-vector with fixed orientation. The finite value of $\Delta$ at any $h$ is compatible with the divergence of the paramagnetic limit at zero temperature and finite angle, which was discussed above. It is illustrated in Fig.~\ref{fig:D(H)_(0,0,kz)}(a) for the order parameter ${\bm \psi}_\mathbf{k} \propto (0, 0, \hat k_z)$ (corresponding to $\eta=0$) with $\Delta(0)=0.81 \Delta_0$. The finite value of $\Delta$ at large $h$ is also illustrated in Fig.~\ref{fig:D(H)_(0,0,kz)}(b) for $\bm{\psi}_\mathbf{k} \propto (\hat k_x,\hat k_y,0)$ (corresponding to $\eta\to\infty$) and an in-plane field, such that $\Delta(0)=0.94 \Delta_0$ and $\Delta(h\gg \Delta_0) \simeq 0.57 \Delta_0^2/h$, and for the isotropic order parameter ${\bm\psi}_\mathbf{k} \propto (\hat k_x,\hat k_y,\hat k_z)$ (corresponding to $\eta=1$) and arbitrary direction of the field, such that $\Delta(0)= \Delta_0$ and $\Delta(h\gg \Delta_0) \simeq 0.81 \Delta_0\sqrt{\Delta_0/h}$.

Figure \ref{fig:D(H)_(0,0,kz)}(a) shows another interesting feature. For an order parameter that allows for a small but finite value of $\bar\theta$, such as illustrated by the purple line in Fig.~\ref{fig:D(H)_(0,0,kz)}(a), the first order transition at $T=0$ takes place between two superconducting phases characterized by different finite values of $\Delta$. At larger $\bar\theta$, the $h$-dependence of $\Delta$ becomes single-valued.

By solving the gap equation at finite temperature and finding the stable solution, which minimizes the free energy \eqref{eq:free_energy_matsubara}, we may now establish the superconducting phase diagram in the $(T,h,\bar\theta)$-phase space for a given symmetry of the order parameter. In particular, a finite temperature tends to make the $h$ dependence of $\Delta$ single valued, see Fig.~\ref{fig:D(H)_(0,0,kz)}(c,d) for $\bm{\psi}_\mathbf{k} \propto (0,0,\hat k_z)$; the transition is always of the second order for ${\bm\psi}_\mathbf{k} \propto (\hat k_x,\hat k_y,0)$ and ${\bm\psi}_\mathbf{k} \propto (\hat k_x,\hat k_y,\hat k_z)$, see Fig~\ref{fig:D(H)_(0,0,kz)}(b). We also checked that no instability towards a spatially modulated Fulde-Ferrell-Larkin-Ovchinnikov (FFLO) \cite{Fulde1964,Larkin1964} phase takes place \cite{SuppMat}.

\begin{figure}
	\centering
	\includegraphics[]{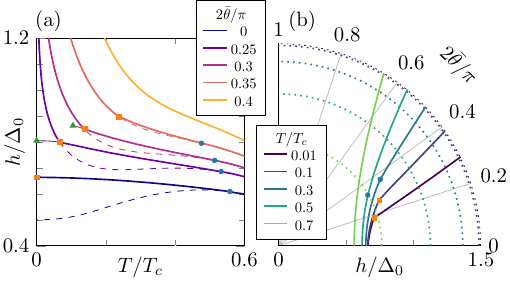}
	\caption{Upper critical field vs.~temperature at different angles (a) and vs.~angle at different temperatures (b) for ${\bm\psi}_\mathbf{k}\propto(0,0,\hat k_z)$ (thick line). Tricritical points are represented by blue circles and orange squares, critical points are represented by green triangles. Panel (a) also shows the upper critical line at the second order transition (dashed line) and the critical line at the first order transition inside the superconducting phase (thin line). Dotted lines in panel (b) are the estimate $H_{c2}^{\rm orb}(T) = H_{c2}^{\rm orb}[1-(T/T_c)^2]$ for the orbital limit assuming a Maki parameter $\alpha=3$.}
	\label{fig:paramagnetic_limit}
\end{figure}

The situation is more complex for the order parameter ${\bm\psi}_\mathbf{k} \propto (0, 0, \hat k_z)$, as illustrated by the $(T,h)$-phase diagram for various angles $\bar\theta$ shown in the left panel of Fig.~\ref{fig:paramagnetic_limit}. As discussed above, the transition from normal to superconducting phase is of the second order both at $T\to 0$ and $T\to T_c$. At small enough $\bar\theta$, two tricritical points appear along the upper critical line and the transition becomes of the first order in-between. Furthermore, a critical line, associated with a first order transition between two superconducting phases, may start from the tricritical point with lower temperature; this critical line may either end at a critical point with finite temperature, or extend down to zero temperature. The transition is of second order for $T>0.56T_c$ or $\bar\theta>0.19\pi$. Our considerations do not include the FFLO phase, which is stabilized at $\bar\theta = 0$ and $T<0.56T_c$. However, at finite $\bar\theta$, we found that the FFLO instability occurs in a narrower temperature range than the first order transition into a uniform phase and, thus, should be less favored \cite{SuppMat}.

\begin{figure}
	\centering
	\includegraphics[]{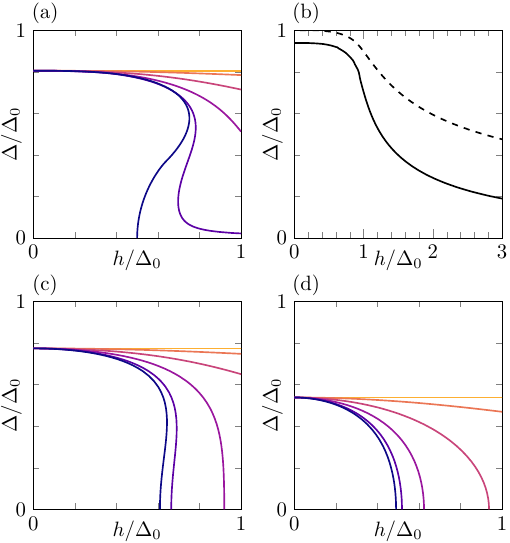}
	\caption{Gap as a function of field. (a, c, d): $\Delta(h)$ for ${\bm\psi}_\mathbf{k} \propto (0,0,\hat k_z)$ and different angles  $\bar\theta$ (same color code as Fig.~\ref{fig:hc(T)_2nd_(0,0,kz)}). (a) $T=0$, (c) $T=0.4 T_c$, (d) $T=0.8T_c$. (b): $\Delta(h)$ at $T=0$ for ${\bm\psi}_\mathbf{k} \propto (\hat k_x,\hat k_y,0)$ and $\bar\theta=\pi/4$, \emph{i.e.} $\mathbf{h}$ in the xy-plane (solid), and ${\bm\psi}_\mathbf{k} \propto (\hat k_x,\hat k_y,\hat k_z)$ and $\bar\theta=0.31\pi$ at arbitrary direction of $\mathbf{h}$ (dashed).}
	\label{fig:D(H)_(0,0,kz)}
\end{figure}

{\it Landau functional.--}
Further insight into the structure of the phase diagram is obtained by expanding the free energy functional \eqref{eq:free_energy_matsubara}, with the gap equation \eqref{eq:gap_triplet_explicit}, in powers of $\Delta$. The expected first order transition requires the analysis of an expansion up to 6$^{\rm th}$ order while the description of a transition between two superconducting phases justifies the following expansion to 8$^{\rm th}$ order in $\Delta$,
\begin{equation}
	F/\nu_0 = a_1 \Delta^2 + a_2 \Delta^4 + a_3 \Delta^6 + a_4 \Delta^8,
	\label{eq:GL}
\end{equation}
in the vicinity of the upper critical line given by Eq.~\eqref{eq:gap_D=0_triplet_psi}, corresponding to $a_1=0$. The coefficients are given as
\begin{eqnarray}
	a_2 &=& \dfrac{ \pi T}{2} \mathrm{Re} \sum_{\omega>0}  \left\langle \frac{|\bm{\psi}_{\mathbf k}|^4}{\omega^2\varepsilon}\Big(\dfrac{\alpha_\mathbf{k}^2}{\varepsilon^2} + \dfrac{ih\sin^4\theta_{\bf k}}{\omega}\Big)\right\rangle_\mathbf{k} 
	\\
	a_3 &=& -\dfrac{ \pi T}{4} \mathrm{Re} \sum_{\omega>0}  \left\langle \frac{|\bm{\psi}_{\mathbf k}|^6}{\omega^3\varepsilon}\Big( \dfrac{\alpha_\mathbf{k}^3}{\varepsilon^4} 
	\right.
	\\
	&&\qquad\qquad\qquad\left.+ \dfrac{ih\alpha_\mathbf{k}\sin^4\theta_{\bf k}}{\omega\varepsilon^2} +\dfrac{ ih \sin^6\theta_{\bf k}}{\omega^2}\Big)\right\rangle_\mathbf{k} 
	 	\nonumber\\
	 a_4 &=& \dfrac{ \pi T}{32}  \mathrm{Re} \sum_{\omega>0} \left\langle \frac{|\bm{\psi}_{\mathbf k}|^4}{\omega^4\epsilon}\Big( \dfrac{5\alpha_\mathbf{k}^4}{\varepsilon^6}  + \dfrac{6ih\alpha_\mathbf{k}^2\sin^4\theta_{\bf k}}{\omega\varepsilon^4}
	\right. 
	  \\ 
	&& \left. +\dfrac{4ih\alpha_\mathbf{k}\sin^6\theta_{\bf k} - h^2\sin^8\theta_{\bf k}}{\omega^2\varepsilon^2} 
	+ \dfrac{5ih\sin^8\theta_{\bf k}}{\omega^3} \Big)\right\rangle_\mathbf{k},\nonumber
\end{eqnarray}
where $\varepsilon = \omega + ih$ and $\alpha_\mathbf{k} = \omega + ih\sin^2\theta_\mathbf{k}$, with $\sin\theta_\mathbf{k} = |\hat{\bm{\psi}}_\mathbf{k}\times\hat{\mathbf{h}}|$ and $\hat{\bm{\psi}}_\mathbf{k} = \bm{\psi}_\mathbf{k}/|\bm{\psi}_\mathbf{k}|$. 

As $a_2>0$ either for $\bar\theta=\pi/2$, or when $h\gg T$, where $a_2 \simeq 7\zeta(3) \langle |\bm{\psi}_{\mathbf k}|^4 \sin^4\theta_\mathbf{k}\rangle_\mathbf{k}/{(4\pi T)^2}$, an expansion of the functional \eqref{eq:GL} up to the $\Delta^4$-term is sufficient to find that the transition remains of the second order. By contrast, at $\bar\theta=0$, $a_2=  {-[\langle|\bm{\psi}_{\mathbf k}|^4\rangle}/{(8\pi T)^2}] \mathrm{Re} \; \psi^{(2)}(1/2+ih/2\pi T)$ changes its sign along the upper critical line, defining the tricritical point $(T_*, h_*)= (0.56T_c, 0.61\Delta_0)$, as in the spin-singlet case. At this point, one can check that $a_3>0$. Thus, only the terms up to $\Delta^6$ need to be retained in the functional \eqref{eq:GL} in order to describe the phase diagram in vicinity of $(T_*, h_*)$. Minimizing the free energy functional, one finds that the upper critical line is given by $a_1=0$ at $a_2>0$, where the transition is of the second order, and by $a_1=a_2^2/4a_3$ at $a_2<0$, where the transition is of the first order.

Taking $\bar\theta$ as an additional parameter, we find a change of sign of the coefficient $a_3$ along the line of tricritical points defined by $a_1=a_2=0$. Indeed, for the model with  ${\bm\psi}_\mathbf{k} \propto (0,0,\hat k_z)$, the condition $a_1=a_2=a_3=0$ is realized at $(T_*, h_*,\bar\theta_*)= (0.27 T_c, 0.91 \Delta_0,0.18\pi)$. At this point, $a_4>0$, thus justifying the expansion of the functional up to order $\Delta^8$ in Eq.~\eqref{eq:free_energy_matsubara}. Its analysis in the vicinity of $(T_*, h_*,\bar\theta_*)$ confirms the structure of the phase diagram illustrated in Fig.~\ref{fig:paramagnetic_limit} in the vicinity of the tricritical point with lower temperature. Namely, in addition to the second order upper critical line defined by $a_1=a_2=0$, $a_3>0$, and the first order upper critical line defined by $a_1=0, 4a_2a_4 = a_3^2$, $a_3<0$, another first order transition line between two competing superconducting phases appears at $16a_1a_4^2 = a_3^3,8a_2a_4 = 3a_3^2$, $a_3 < 0$ \cite{SuppMat}.

{\it Discussion.--} The striking divergence of the paramagnetic limit at $T\to 0$ is cut off by the orbital effect of the magnetic field, which we neglected so far. It is known that the orbital critical field is always dominant close to $T_c$.  While the paramagnetic limit may become relevant at intermediate temperature, it becomes ineffective again at $T\to 0$ due to its divergence.
As a consequence, the temperature dependence of the upper critical field anisotropy provides interesting information about the spin structure of the order parameter. 

Among candidate materials for triplet superconductivity, UPt$_3$ is a rare example where multiple superconducting phases and the anisotropy of the upper critical line in directions parallel and perpendicular to the basal plane provide strong evidence of a triplet order parameter~\cite{Joynt2002}. In particular, the weaker temperature dependence of the perpendicular upper critical field at low $T$ hints to a triplet order parameter aligned with the $\hat{\mathbf c}$-axis~\cite{Choi1991}. This, however, contradicts the absence of reduction of the Knight shift~\cite{Tou1998}, and the structure of the order parameter remains unsettled, see, e.g., \cite{Nomoto2016}. To confirm the prediction of a $d$-vector pinned to the $\hat{\mathbf c}$-axis, our results suggest investigating the temperature dependence of the upper critical field anisotropy at lower temperatures than those investigated in \cite{Keller1994,Kittaka2013}. In particular, Eq.~\eqref{eq:hc_divergence} predicts that, at low enough temperature, the paramagnetic limit is ineffective, and the upper critical field is orbitally limited, as soon as its orientation deviates from $\hat{\mathbf c}$-axis by an angle $\sim \sqrt{\ln \alpha/\ln(T_c/T)}$, where $\alpha= \sqrt{2}H^{\rm orb}_{c2}/H_P$ is the so-called Maki parameter that measures the ratio between the orbital limit, $H^{\rm orb}_{c2}\sim \Phi_0/\xi^2$, and the Pauli limit, $H_P$, provided that $\alpha>1$ \cite{Maki1964b}. Here $\Phi_0$ is the superconducting flux quantum and $\xi$ is the superconducting coherence length. A comparison between the paramagnetic and orbital limits is shown Fig \ref{fig:paramagnetic_limit}(b). Thus, our model predicts that the magnetic field anisotropy strongly depends on the temperature, which is in marked contrast with conventional theories that are based on the anisotropy of the Fermi surface, yielding a weak temperature dependence of the critical field anisotropy \cite{Tinkham1996}.

\begin{acknowledgments}
We thank J.-P.~Brison and K.~Hasselbach for fruitful discussions. We acknowledge the support of France 2030 ANR QuantForm-UGA and ANR-21-CE30-0035 (TRIPRES).
\end{acknowledgments}

\bibliographystyle{apsrev4-1}
\bibliography{Bibliography}

\end{document}


\title{Supplemental Material for ``Paramagnetic limit of spin-triplet superconductors"}

\maketitle

This supplemental material contains technical details on the derivation and analysis of the paramagnetic limit of spin-triplet superconductors that were omitted in the main text. In Sec.~\ref{se:free_energy}, we detail the process to obtain the free energy difference as a sum over Matsubara frequencies Eq.~(4). In Sec.~\ref{se:asymptotic}, we derive the asymptotic value of the gap $\Delta(h)$ for $h\gg\Delta$ at $T=0$ Eq.~(9). In Sec.~\ref{se:FFLO}, we investigate the possibility of a spatially modulated Fulde-Ferrell-Larkin-Ovchinnikov (FFLO) phase and show that it may be stabilized in part of the region where we found a first order transition at low $T$, $\bar\theta$. In Sec \ref{se:tricritical}, we derive the conditions on the coefficients of the generalized Ginzburg-Landau functional given in Eq.~(11) of the main text that define the two tricritical lines and the critical line of its phase diagram. 

\section{Free energy}
\label{se:free_energy}

To derive Eq.~(4) in the main text, we insert the identity
\begin{equation}
	\cosh^2\frac{x}{2} = \prod_{n=-\infty}^{+\infty} \left[ 1 + \dfrac{x^2}{\pi^2(2n+1)^2} \right] 
	\label{eq:cosh2}
\end{equation}
into Eq.~(1) of the main text. After taking the continuum limit $\sum_\mathbf{k} \dots \rightarrow \nu_0 \int d\xi_\mathbf{k} \langle \dots \rangle_\mathbf{k}$ where $\nu_0$ is the density of states at the Fermi level, one obtains
\begin{equation}
	F_S = \nu_0 \int_{|\xi_{\mathbf k}|<{\cal E}} d\xi_\mathbf{k} \left\langle \xi_\mathbf{k} + \dfrac{T}{2} \sum_{\pm,\omega} \ln\left[\dfrac{\omega^2}{4(\omega^2 + E_\mathbf{k,\pm}^2)}\right] \right\rangle_\mathbf{k} + \dfrac{\nu_0 |\Delta|^2}{\lambda},
	\label{eq:free_energy_intermediary}
\end{equation}
where $\omega = \pi T (2n+1)$ with $n$ integer. We subtract the normal contribution $F_N = F_S(\Delta=0)$ to obtain the difference $F = F_S - F_N$:
\begin{equation}
	F = \dfrac{\nu_0T}{2} \sum_{\pm,|\omega|<\mathcal{E}} \left\langle \int d\xi_{\mathbf k} \ln\left[\dfrac{\omega^2 + (\xi_\mathbf{k} \pm h)^2}{\omega^2 + \xi_\mathbf{k}^2 + \Delta^2|\boldsymbol{\psi}_\mathbf{k}|^2 + h^2 \pm 2|h|(\xi_\mathbf{k}^2 + \Delta^2|\boldsymbol{\psi}_\mathbf{k}\cdot\hat{\mathbf{h}}|^2)^{1/2}}\right] \right\rangle_\mathbf{k} + \dfrac{\nu_0 |\Delta|^2}{\lambda},
\end{equation}
where the integral and the sum have been exchanged. The integral converges and can be taken to infinity. However, the sum now diverges and must be cut off at the relevant energy scale for the pairing $\cal E$. The integral can be computed analytically and leads to Eq.~(4) in the main text.

\section{Asymptotic value of $\Delta(h)$ at $h\gg\Delta$}
\label{se:asymptotic}

We compute the asymptotic value of the gap at large $h$ and $T=0$, Eq.~(9) in the main text, starting from the general gap equation Eq.~(5) in the main text, reading at $T=0$,
\begin{equation}
	\dfrac{1}{\lambda} = \int_{0}^{\cal E} d\omega \left\langle |\bm{\psi}_{\mathbf k}|^2\textrm{Re} \; \Omega_\mathbf{k}^{-1} + \dfrac{|\bm{\psi}_{\mathbf k}\times\mathbf{h}|^2}{|\Omega_\mathbf{k}|^2 \textrm{Re} \; \Omega_\mathbf{k}} \right\rangle_\mathbf{k}.
	\label{eq:gap_triplet_T=0}
\end{equation}
The second term of Eq.~\eqref{eq:gap_triplet_T=0} converges at large $\omega$, while the divergence of the first term must be cut off. The cutoff energy can be taken to infinity after the first term is regularized by subtracting its value at $h=0$
\begin{equation}
	\dfrac{1}{\lambda} = \int_0^{\cal E} d\omega \left\langle \dfrac{|\boldsymbol{\psi}_\mathbf{k}|^2}{\sqrt{\omega^2+\Delta(0)^2|\boldsymbol{\psi}_\mathbf{k}|^2}} \right\rangle_\mathbf{k} = \left\langle|\boldsymbol{\psi}_\mathbf{k}|^2\ln\dfrac{2\cal E}{\Delta(0)|\boldsymbol{\psi}_\mathbf{k}|}\right\rangle_\mathbf{k},
\end{equation}
which defines the value of the gap at $T=h=0$, $\Delta(0) = \Delta_0 \exp \langle -|\bm{\psi}_{\mathbf k}|^2\ln|\bm{\psi}_{\mathbf k}|\rangle_\mathbf{k}$, with $\Delta_0 = 2\mathcal{E} e^{-1/\lambda} \approx 1.76 T_c$ the BCS gap. With that definition, we have $\langle|\boldsymbol{\psi}_\mathbf{k}|^2\ln[2\mathcal{E}/\Delta(0)|\boldsymbol{\psi}_\mathbf{k}|]\rangle_\mathbf{k} = \ln(2\mathcal{E}/\Delta_0)$. To enforce the asymptotic condition $h\gg\Delta$, we keep $\Delta$ only where it is necessary for convergence. Putting $\Delta=0$ in the first term of Eq.~(5) reduces it to $\ln(\mathcal{E}/h)$. However, the second term of Eq.~(5) diverges as $1/\omega$ at low $\omega$ if $\Delta=0$ due to $\textrm{Re} \; \Omega_\mathbf{k}$, and an expansion is needed. Therefore, one needs to keep $\Delta$ finite in $\textrm{Re} \; \Omega_\mathbf{k}$, which at  $h \gg \Delta, \omega$, can be approximated as  $\textrm{Re} \;\Omega_\mathbf{k} \approx  (\omega^2 + \Delta^2|\bm{\psi}_\mathbf{k}\times\hat{\mathbf{h}}|^2)^{1/2}$. The approximated gap equation therefore reads:
\begin{equation}
	\ln\dfrac{2h}{\Delta_0} = \left\langle \int_0^\infty d\omega \dfrac{h^2|\boldsymbol{\psi_k}\times\hat{\mathbf{h}}|^2}{(\omega^2+h^2)\sqrt{\omega^2+\Delta^2|\bm{\psi}_\mathbf{k}\times\hat{\mathbf{h}}|^2}} \right\rangle_\mathbf{k}.
\end{equation}
The integral can be computed analytically and leads to
\begin{equation}
	\dfrac{h|\boldsymbol{\psi}_\mathbf{k}\times\hat{\mathbf{h}}|^2}{\sqrt{h^2-\Delta^2|\bm{\psi}_\mathbf{k}\times\hat{\mathbf{h}}|^2}} \arccos\left(\dfrac{h}{\Delta|\bm{\psi}_\mathbf{k}\times\hat{\mathbf{h}}|}\right) \underset{h\gg\Delta}{\simeq} |\boldsymbol{\psi}_\mathbf{k}\times\hat{\mathbf{h}}|^2 \ln\left(\dfrac{2h}{\Delta|\bm{\psi}_\mathbf{k}\times\hat{\mathbf{h}}|}\right).
\end{equation}
The effective angle $\bar\theta$ appears in the averaged squared cross-product $\langle|\boldsymbol{\psi}_\mathbf{k}\times\hat{\mathbf{h}}|^2\rangle_\mathbf{k} = 1 - \langle |\boldsymbol{\psi}_\mathbf{k}\cdot\hat{\mathbf{h}}|^2 \rangle_\mathbf{k} = \sin^2\bar\theta$. Thus, the approximated gap equation becomes 
\begin{equation}
	\sin^2\bar\theta \ln(2h/\Delta) = \ln(2h/\Delta_0) + \langle |\bm{\psi}_{\mathbf k}\times\hat{\mathbf{h}}|^2\ln|\bm{\psi}_{\mathbf k}\times\hat{\mathbf{h}}| \rangle_\mathbf{k},
\end{equation}
which is rewritten as Eq.~(9) in the main text.

\section{Stability of FFLO phase}
\label{se:FFLO}

We examine the possibility of a second order transition from the normal state to a phase with spatially modulated gap $\Delta(\mathbf{x})= \Delta_\mathbf{q} e^{i\mathbf{q}\cdot\mathbf{x}}$. Assuming the order parameter to be small at the transition, we start from the linearized inhomogeneous Eilenberger equation
\begin{equation}
	iv_F \; \hat{\mathbf{k}}\cdot\mathbf{q} - \{ \omega + i\mathbf{h}\cdot\boldsymbol{\sigma}, \hat f_{\mathbf{k},\omega}(\mathbf{q})\} = 2\Delta\boldsymbol{\psi}_\mathbf{k}\cdot\boldsymbol{\sigma},
	\label{eq:Eilenberger_linear}
\end{equation}
where $v_F$ is the Fermi velocity, $\hat{\mathbf{k}}$ is the normalized momentum and $\hat f_{\mathbf{k},\omega}(\mathbf{q})$ is the anomalous quasiclassical Green function matrix in spin space obeying the gap equation
\begin{equation}
	\Delta_\mathbf{k}(\mathbf{q}) = \dfrac{\lambda \pi T}{2} \sum_{|\omega|<\cal E}  \mathrm{Tr}\langle\boldsymbol{\psi}_\mathbf{k}^*\cdot\boldsymbol{\sigma} \hat f_{\mathbf{k},\omega}(\mathbf{q})\rangle_\mathbf{k}.
	\label{eq:gap_f}
\end{equation}
Equations \eqref{eq:Eilenberger_linear} and \eqref{eq:gap_f} lead to an implicit equation for the second order critical field of the FFLO state. After summation over Matsubara frequencies and regularization, it reads:
\begin{equation}
	\ln\dfrac{T}{T_c} = \left\langle \psi\left(\dfrac{1}{2}\right) - |\boldsymbol{\psi}_\mathbf{k}\times\hat{\mathbf{h}}|^2 \mathrm{Re} \; \psi\left(\dfrac{1}{2}+i\dfrac{v_F \:\hat{\mathbf{k}}\cdot\mathbf{q}}{4\pi T} \right) -  \dfrac{|\boldsymbol{\psi}_\mathbf{k}\cdot\hat{\mathbf{h}}|^2}{2} \sum_\pm \psi\left(\dfrac{1}{2}+i\dfrac{v_F \: \hat{\mathbf{k}}\cdot\mathbf{q} \pm 2h}{4\pi T}\right) \right\rangle_\mathbf{k},
	\label{eq:gap_2nd_modulated}
\end{equation}
where $\psi$ is the digamma function and $\hat{\mathbf{h}} = \mathbf{h}/|\mathbf{h}|$. Assuming a small modulation at the second order transition $v_F|\mathbf{q}| \ll \Delta_0$, we expand this expression at small $|\mathbf{q}|$, leading to:
\begin{equation}
	a_1 + \dfrac{1}{2}\left(\dfrac{v_F}{4\pi T}\right)^2 |\mathbf{q}|^2 b_2  = 0,
	\label{eq:expansion_q2}
\end{equation}
where $a_1$ was defined in the main text as the coefficient in front of $\Delta^2$ in the free energy functional, and $(v_F/4\pi T)^2|\mathbf{q}|^2b_2$ is the correction of the same functional due to the modulation with
\begin{equation}
	b_2 = -\left\langle (\hat{\mathbf{k}}\cdot\hat{\mathbf{q}})^2\left[|\boldsymbol{\psi}_\mathbf{k}\times\hat{\mathbf{h}}|^2 \psi^{(2)}\left(\dfrac{1}{2}\right) + |\boldsymbol{\psi}_\mathbf{k}\cdot\hat{\mathbf{h}}|^2 \mathrm{Re} \; \psi^{(2)}\left(\dfrac{1}{2}+\dfrac{ih}{2\pi T}\right) \right]\right\rangle_\mathbf{k}.
	\label{eq:b2}
\end{equation}
Here, $\psi^{(2)}$ is the second order derivative of the digamma function and $\hat{\mathbf{q}} = \mathbf{q}/|\mathbf{q}|$.

The sign of $b_2$ determines  whether an FFLO state is favored compared to the uniform state: namely, if it is negative (positive), a finite modulation vector is (un)favored. At $\bar\theta = 0$, $b_2\propto a_2$, i.e., the FFLO state is favored in the same temperature range as the first order transition, $T\in[0, 0.56T_c]$. 

\begin{figure}
	\centering
	\includegraphics[]{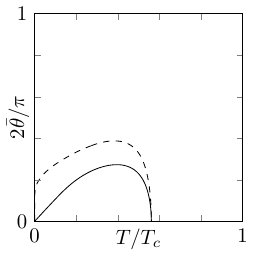}
	\caption{Projection onto the $(T,\; \bar\theta)$ of the FFLO critical points along the second order transition critical field between normal and superconducting phases (solid) and of tricritical points along the upper critical line between	normal and superconducting phases (dashed) for an order parameter $\mathbf{d_k} \propto (0,0,\hat k_z)$. The solid line corresponds to $b_2=0$ and the region below to $b_2 < 0$.}
	\label{fig:FFLO}
\end{figure}

For an order parameter with fixed orientation, the expression for $b_2$ simplifies to
 \begin{equation}
	b_2 = \left\langle (\hat{\mathbf{k}}\cdot\hat{\mathbf{q}})^2|\boldsymbol{\psi}_\mathbf{k}|^2\right\rangle_\mathbf{k}\left[\sin^2\bar\theta\,|\psi^{(2)}\left(\dfrac{1}{2}\right)\!|- \cos^2\bar\theta\mathrm{Re} \; \psi^{(2)}\left(\dfrac{1}{2}+\dfrac{ih}{2\pi T}\right) \right].
	\label{eq:b2_fixed}
\end{equation}
Thus the sign of $b_2$ does not depend on the direction of $\hat{\mathbf{q}}$, but is determined solely by the angle $\bar\theta$. In particular, as $h/T$ spans the entire range from $0$ to $\infty$  when $T$ decreases from $T_c$ to zero, $b_2$ is positive at all temperatures for
\begin{equation}
\tan^2\bar\theta>\frac{{\rm max}_x\mathrm{Re} \; \psi^{(2)}\left(\dfrac{1}{2}+ix\right)}{|\psi^{(2)}\left(\dfrac{1}{2}\right)\!|}\approx 0.21.
\end{equation}
Thus, at $\bar\theta>0.14\pi$, the second order transition into a FFLO state is no longer favored at any temperature. Note that this angle is smaller than the one, where the first order transition disappears. Furthermore, at angles $0<\bar\theta<0.14\pi$, the temperature range where an FFLO modulated phases is favored is smaller than the temperature range where the transition is first order as shown in Fig. \ref{fig:FFLO}. 

If the orientation of $\bm{\psi}_\mathbf{k}$ varies in momentum space, the sign of $b_2$ depends on the direction of $\hat{\mathbf{q}}$. In that case, a temperature range exists where the FFLO modulated phase is favored, if
\begin{equation}
	\dfrac{\left\langle (\hat{\mathbf{k}}\cdot\hat{\mathbf{q}})^2|\boldsymbol{\psi}_\mathbf{k}\times\hat{\mathbf{h}}|^2 \right\rangle_\mathbf{k}}{\left\langle(\hat{\mathbf{k}}\cdot\hat{\mathbf{q}})^2|\boldsymbol{\psi}_\mathbf{k}\cdot\hat{\mathbf{h}}|^2\right\rangle_\mathbf{k}} <\frac{{\rm max}_x\mathrm{Re} \; \psi^{(2)}\left(\dfrac{1}{2}+ix\right)}{|\psi^{(2)}\left(\dfrac{1}{2}\right)\!|}
\end{equation}
for some $\hat{\mathbf{q}}$. With this condition, we can show, in particular, that for the order parameters ${\bm\psi}_\mathbf{k} \propto (\hat k_x,\hat k_y,0)$ and ${\bm\psi}_\mathbf{k} \propto (\hat k_x,\hat k_y,\hat k_z)$ discussed in the main text, no FFLO modulated phase can be stabilized. 
 
\section{Critical and tricritical lines of a generalized Ginzburg-Landau functional}
\label{se:tricritical}
\begin{figure}
	\centering
	\includegraphics[]{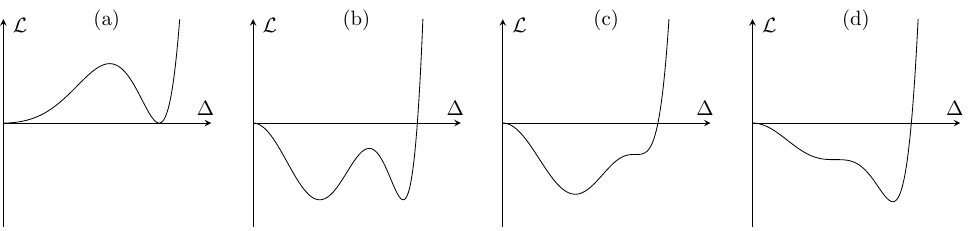}
	\caption{$\mathcal{L}(\Delta)$ at the first order transition (a), transition between two superconducting phases (b), appearance of a second metastable superconducting phase in addition to the stable superconducting phase (c, d).}
	\label{fig:L(Delta)}
\end{figure}

At the second order transition between normal and superconducting phases, $\Delta$ vanishes. Thus, in the vicinity of the transition an expansion of the free energy in powers of $\Delta^2$ is possible, and the first few coefficients of this expansion capture its general behavior. The minimal order necessary to account for two possible minima at finite $\Delta$ and study the transition between the two is $8$. Hence we study the Landau functional ${\cal L}\propto F$ given as
\begin{equation}
	\mathcal{L} = a_1 \Delta^2 + a_2 \Delta^4 + a_3 \Delta^6 + a_4 \Delta^8,
\end{equation}
where $a_4>0$ to ensure stability. At $a_4$ fixed, the phase diagram of $\mathcal{L}$ is characterized by critical and tricritical lines delimiting three critical surfaces (second order transition between normal and superconducting phases, first order transition between normal and superconducting phases, first order between two superconducting phases) that separate possible phases in $(a_1,a_2,a_3)$ space. Each line is described by conditions on the extrema and thus on the coefficients. The extrema of $\mathcal{L}$ are either at $x=0$ or at the real strictly positive solutions $x = a_4^{1/4}\Delta^2$ of:
\begin{eqnarray}
	a_1 + 2a_2 x + 3a_3x^2 + 4x^3 = 0,
	\label{eq:dL|dD=0}
\end{eqnarray}
where we rescaled the coefficient as $a_1 \to a_1a_4^{1/4}$, $a_2 \to a_2a_4^{1/2}$, $a_3 \to a_3a_4^{3/4}$. The region $a_1,a_2,a_3 > 0$ is a normal phase. In that region, $\mathcal{L}$ has a single minimum at $\Delta=0$. The second order transition from the normal state occurs on the surface $a_1 = 0$, where the extremum at $\Delta=0$ changes from a minimum to a maximum. We determine now its intersection with the first order transition surface. The first order transition from the normal state  occurs when $\mathcal{L}(\Delta)=0$ at a minimum with $\Delta\neq0$ as illustrated Fig.~\ref{fig:L(Delta)}(a). We thus solve:
\begin{subequations}
	\begin{eqnarray}
	a_1 + 2a_2x + 3a_3x^2 + 4x^3 &=& 0, \\
	a_1  + a_2x + a_3x^2 + x^3 &=& 0.
	\end{eqnarray}
\end{subequations}
Subtracting the two equations to eliminate $a_1$ and assuming $x\neq 0$, we obtain a quadratic equation that can be solved to yield
\begin{equation}
x = \dfrac{1}{3}\left(-a_3 \pm \sqrt{a_3^2 - 3a_2}\right).
\end{equation}
Again subtracting the two equations, this time to eliminate cubic terms, and substituting the above solutions for $x$, we obtain an equation relating the three coefficients,
\begin{equation}
	a_{1}^{(\pm)} = \dfrac{1}{27}\left[-a_3(2a_3^2-9a_2) \pm 2(a_3^2-3a_2)^{3/2}\right].
	\label{eq:a1_first}
\end{equation}
The conditions $x,a_1>0$ to obtain a first order transition impose constraints on $a_2$ and $a_3$. Namely, at $a_3>0$, the condition $x>0$ requires $a_2<0$. This discards $a_{1}^{(-)}$ which is positive only for positive $a_2\in[a_3^2/4,a_3^2/3]$. Thereby, $a_{1}^{(+)}=0$ yields a critical line,  
\begin{equation}
a_1=a_2=0, \qquad a_3>0
\label{eq-l1}
\end{equation}
where the transition changes form first to second order.
By contrast, at $a_3<0$, the coefficient $a_2$ may be positive. The solution $a_{1}^{(-)}$ is always negative and can thus be discarded again. Setting $a_{1}^{(+)}=0$ in Eq.~\eqref{eq:a1_first} then yields another critical line
\begin{equation}
a_1=0,\quad a_2=a_3^2/4, \quad a_3<0.
\label{eq-l2}
\end{equation}

Equivalent expressions where derived in \cite{Casalbuoni2004} to address the existence of two tricritical lines at the transition between normal and two types of modulated FFLO phases.

A first order transition between two superconducting phases occurs when two minima of ${\cal L}(\Delta)$ at $\Delta\neq0$ become degenerate as illustrated in Fig.~\ref{fig:L(Delta)}(b). This may happen only if $a_1,a_3 < 0$ and $a_2 > 0$. Namely, two such minima may only exist when the discriminant of the cubic equation  \eqref{eq:dL|dD=0},
\begin{eqnarray}
D = \frac1{64}\left(108a_1a_2a_3 - 32a_2^3 + 9a_2^2a_3^2 - 27a_1a_3^3 - 108 a_1^2\right),
\end{eqnarray}
is positive. The discriminant changes sign at
\begin{eqnarray}
	a_1^{(\pm)} = \dfrac{a_3}{2} \left( a_2 - \dfrac{a_3^2}{4} \right) \pm \left( \dfrac{a_3^2}{4} - \dfrac{2a_2}{3} \right)^{3/2}.
	\label{eq:a1_vardelta=0}	
\end{eqnarray}
This yields a range of $a_1^{(-)}<a_1<a_1^{(+)}$, where two minima exist if $a_1<0$ and, thus, a transition between two superconducting phases may occur. The cases where $a_1=a_1^{(\pm)}$ such that an additional (metastable) minimum appears are illustrated in Figs.~\ref{fig:L(Delta)}(c) and \ref{fig:L(Delta)}(d). This range shrinks to zero when $a_2=3a_3^2/8$ yielding $a_1^{(\pm)}=a_3^3/16$. This defines the third critical line at $a_3<0$, i.e.,
\begin{equation}
a_1=a_3^3/16,\quad a_2=3a_3^2/8, \quad a_3<0.
\label{eq-l3}
\end{equation}
To summarize, we obtain three critical lines given by Eqs.~\eqref{eq-l1}, \eqref{eq-l2}, and \eqref{eq-l3} that meet at the point $a_1=a_2=a_3$ as shown in Fig.~4 of the main text. The coefficients may be rescaled to obtain the tricritical and critical lines given below Eqs.~(11)-(14) in the main text.

\bibliography{Bibliography}
\bibliographystyle{apsrev4-1}